\documentclass{article}

\usepackage{color}
\usepackage{amsmath, amssymb}

\newcommand{\ZM}{\mathbb{Z}}

\newcommand{\CM}{\mathbb{C}}
\newcommand{\PM}{\mathbb{P}}

\newcommand{\bec}[1]{\mbox{\boldmath $#1$}}
\newcommand{\miniket}[1]{\vert#1\rangle}

\begin{document}

\title{{\bf Absolute zeta functions for \\ 
zeta functions of quantum cellular automata}
\vspace{15mm}}

\author{Jir\^o AKAHORI, $\quad$ Norio KONNO$^{\ast}$ \\ \\
Department of Mathematical Sciences \\
College of Science and Engineering \\
Ritsumeikan University \\
1-1-1 Noji-higashi, Kusatsu, 525-8577, JAPAN \\
e-mail: akahori@se.ritsumei.ac.jp, \  n-konno@fc.ritsumei.ac.jp$^{\ast}$ \\
\\
\\
Iwao SATO \\ \\
Oyama National College of Technology \\
771 Nakakuki, Oyama 323-0806, JAPAN \\
e-mail: isato@oyama-ct.ac.jp 
}

\date{\empty }

\maketitle

\vspace{50mm}


\vspace{20mm}


\begin{small}
\par\noindent
\\
\end{small}









\clearpage

\begin{abstract}
Our previous work dealt with the zeta function for the interacting particle system (IPS) including quantum cellular automaton (QCA) as a typical model in the study of ``IPS/Zeta Correspondence". On the other hand, the absolute zeta function is a zeta function over $\mathbb{F}_1$ defined by a function satisfying an absolute automorphy. This paper proves that a new zeta function given by QCA is an absolute automorphic form of weight depending on the size of the configuration space. As an example, we calculate an absolute zeta function for a tensor-type QCA, and show that it is expressed as the multiple gamma function. In addition, we obtain its functional equation by the multiple sine function.
\end{abstract}

\vspace{10mm}

\begin{small}
\end{small}

\vspace{10mm}

\section{Introduction \label{sec01}}
In our previous papers \cite{KomatsuEtAl2022, KiumiEtAl2022}, we investigated zeta functions for the interacting particle systems (IPSs) including quantum cellular automata (QCA) and probabilistic cellular automata (PCA) as typical models in the study of ``IPS/Zeta Correspondence". Concerning IPS, see \cite{Durett1988}, for example. Here we focus on two-state QCA on $\mathbb{P}_N = \{ 0,1, \ldots, N-1 \}$, where $\mathbb{P}_N$ denotes the path space with $N$ sites. There are two states ``0" or ``1" at each site. Let $\eta (x) \in \{0,1\}$ be the state of the site $x \in \PM_N$, i.e., $x=0,1,\ldots, N-1$. The configuration space is $\{0,1\}^{\mathbb{P}_N}$ with $2^N$ elements. 

On the other hand, the absolute zeta function is a zeta function over $\mathbb{F}_1$ defined by a function satisfying an absolute automorphy. Here $\mathbb{F}_1$ can be viewed as a kind of limit of $\mathbb{F}_p$ as $p \to 1$, where $\mathbb{F}_p = \mathbb{Z}/p \mathbb{Z}$ stands for the field of $p$ elements for a prime number $p$. As for absolute zeta function, see \cite{CC, KF1, Kurokawa3, Kurokawa, KO, KT3, KT4, Soule}.

In this setting, we introduce our zeta function $\zeta_N (u)$ determined by a $Q^{(g)}_N$ which is a $2^N \times 2^N$ time evolution matrix of QCA on $\mathbb{P}_N$. More precisely, $\zeta_N (u) = \det ( I_{2^N} - u Q^{(g)}_N )^{-1}$, where $I_n$ is the $n \times n$ identity matrix. We show that if $Q^{(g)}_N$ is an orthogonal matrix, then $\zeta_N (u)$ is an absolute automorphic form of weight $- 2^N$. After that, we consider an absolute zeta function $\zeta_{\zeta_N} (s)$ for our zeta function $\zeta_N (u)$. As an example, we calculate $\zeta_{\zeta_N} (s)$ for a tensor-type QCA, and prove that it is expressed as the multiple gamma function via the multiple Hurwitz zeta function. In addition, we obtain the functional equation for $\zeta_{\zeta_N} (s)$ by the multiple sine function. The present manuscript is the first step of the study on a relationship between the QCA and the absolute zeta function. As for the absolute zeta function for a zeta function based on the quantum walk, see \cite{Konno2023}.

The rest of this paper is organized as follows. Section \ref{sec02} briefly describes the absolute zeta function and its related topics. In addition, we give four examples which will be used in Section \ref{sec05}. Section \ref{sec03} deals with the definition of the QCA. In Section \ref{sec04}, we treat our zeta function defined by QCA. In Section \ref{sec05}, we calculate the absolute zeta function for a tensor-type QCA. Finally, Section \ref{sec06} is devoted to summary.

\section{Absolute Zeta Functions \label{sec02}}
First we introduce the following notation: $\mathbb{Z}$ is the set of integers, $\mathbb{Z}_{>} = \{1,2,3, \ldots \}$,  $\mathbb{R}$ is the set of real numbers, and $\mathbb{C}$ is the set of complex numbers. 

In this section, we briefly review the framework on the absolute zeta functions, which can be considered as zeta function over $\mathbb{F}_1$, and absolute automorphic forms (see \cite{Kurokawa3, Kurokawa, KO, KT3, KT4} and references therein, for example). 

Let $f(x)$ be a function $f : \mathbb{R} \to \mathbb{C} \cup \{ \infty \}$. We say that $f$ is an {\em absolute automorphic form} of weight $D$ if $f$ satisfies
\begin{align*}
f \left( \frac{1}{x} \right) = C x^{-D} f(x)
\end{align*}
with $C \in \{ -1, 1 \}$ and $D \in \mathbb{Z}$. The {\em absolute Hurwitz zeta function} $Z_f (w,s)$ is defined by
\begin{align*}
Z_f (w,s) = \frac{1}{\Gamma (w)} \int_{1}^{\infty} f(x) \ x^{-s-1} \left( \log x \right)^{w-1} dx,
\end{align*}
where $\Gamma (x)$ is the gamma function (see \cite{Andrews1999}, for instance). Then taking $u=e^t$, we see that $Z_f (w,s)$ can be rewritten as the Mellin transform: 
\begin{align}
Z_f (w,s) = \frac{1}{\Gamma (w)} \int_{0}^{\infty} f(e^t) \ e^{-st} \ t^{w-1} dt.
\label{kirishima01}
\end{align}
Moreover, the {\em absolute zeta function} $\zeta_f (s)$ is defined by 
\begin{align*}
\zeta_f (s) = \exp \left( \frac{\partial}{\partial w} Z_f (w,s) \Big|_{w=0} \right).
\end{align*}
Here we introduce the {\em multiple Hurwitz zeta function of order $r$}, $\zeta_r (s, x, (\omega_1, \ldots, \omega_r))$, the {\em multiple gamma function of order $r$}, $\Gamma_r (x, (\omega_1, \ldots, \omega_r))$, and the {\em multiple sine function of order $r$}, $S_r (x, (\omega_1, \ldots, \omega_r))$, respectively (see \cite{Kurokawa3, Kurokawa, KT3}, for example): 
\begin{align}
\zeta_r (s, x, (\omega_1, \ldots, \omega_r))
&= \sum_{n_1=0}^{\infty} \cdots \sum_{n_r=0}^{\infty} \left( n_1 \omega_1 + \cdots + n_r \omega_r + x \right)^{-s}, 
\label{kirishima03}
\\
\Gamma_r (x, (\omega_1, \ldots, \omega_r)) 
&= \exp \left( \frac{\partial}{\partial s} \zeta_r (s, x, (\omega_1, \ldots, \omega_r)) \Big|_{s=0} \right),
\label{kirishima04}
\\
S_r (x, (\omega_1, \ldots, \omega_r))
&= \Gamma_r (x, (\omega_1, \ldots, \omega_r))^{-1} \ \Gamma_r (\omega_1+ \cdots + \omega_r - x, (\omega_1, \ldots, \omega_r))^{(-1)^r}.
\label{kirishima04b}
\end{align}
\par
Now we present the following key result derived from Theorem 4.2 and its proof in Korokawa \cite{Kurokawa} (see also Theorem 1 in Kurokawa and Tanaka \cite{KT3}):

\vspace*{12pt}
\noindent
{\bf Theorem~1:}
For $\ell \in \mathbb{Z}, \ m(i) \in \mathbb{Z}_{>} \ (i=1, \ldots, a), \ n(j) \in \mathbb{Z}_{>} \ (j=1, \ldots, b)$, put 
\begin{align*}
f(x) = x^{\ell/2} \ \frac{\left( x^{m(1)} - 1 \right) \cdots \left( x^{m(a)} - 1 \right)}{\left( x^{n(1)} - 1 \right) \cdots \left( x^{n(b)} - 1 \right)}.
\end{align*}
Then we have 
\begin{align}
Z_f (w, s) 
&= \sum_{I \subset \{1, \ldots, a \}} (-1)^{|I|} \ \zeta_b \left( w, s - {\rm deg} (f) + m \left( I \right), \bec{n} \right),
\label{mkusatsu01}
\\
\zeta_f (s) 
&= \prod_{I \subset \{1, \ldots, a \}} \Gamma_b \left( s - {\rm deg} (f) + m \left( I \right), \bec{n} \right)^{ (-1)^{|I|}},
\label{mkusatsu02}
\\
\zeta_f \left( D-s \right)^C 
&= \varepsilon_f (s) \ \zeta_f (s),
\label{mkusatsufe}
\end{align}
where
\begin{align*}
|I| 
&= \sum_{i \in I} 1, \quad {\rm deg} (f) = \frac{\ell}{2} + \sum_{i=1}^a m(i)- \sum_{j=1}^b n(j), \quad m \left( I \right) = \sum_{i \in I} m(i),
\\
\bec{n} 
&= \left( n(1), \ldots, n(b) \right), \quad D = \ell + \sum_{i=1}^a m(i)- \sum_{j=1}^b n(j), \quad C=(-1)^{a-b}, 
\\
\varepsilon_f (s) 
&= \prod_{I \subset \{1, \ldots, a \}} S_b \left( s - {\rm deg} (f) + m \left( I \right), \bec{n} \right)^{ (-1)^{|I|}}.
\end{align*}

\vspace*{12pt}
\noindent
We should note that Eq. \eqref{mkusatsufe} is called the {\em functional equation}. 
\par
From now on, we give four examples of $f(x)$ which will be discussed in detail in Section \ref{sec05}. 
\par
\
\par
Case (i). 
\par
\begin{align}
f_1 (x) = \frac{1}{(x -1)^2}.
\label{kirishima02}
\end{align}
Then we see
\begin{align*}
f_1 \left( \frac{1}{x} \right) = x^{2} f_1(x),
\end{align*}
thus $f_1$ is an absolute automorphic form of weight $- 2 = -2^1$. Noting that $\ell =0, \ a=0, \ b=2, \ n(1)=n(2)=1, \ {\rm deg} (f_1) = D = -2, \ \bec{n} =(1,1), \  C=1,$ and $\varepsilon_{f_1} (s) = S_2 \left( s + 2, (1,1) \right)$, it follows from Theorem 1 that
\begin{align*}
Z_{f_1} (w, s) 
&= \zeta_2 \left( w, s + 2, (1,1) \right),
\\
\zeta_{f_1} (s)
&= \Gamma_2 \left( s + 2, (1,1) \right),
\\
\zeta_{f_1} (-2-s)
&= S_2 \left( s + 2, (1,1) \right) \ \zeta_{f_1} (s).
\end{align*}
So $Z_{f_1} (w,s)$ and $\zeta_{f_1} (s)$ can be obtained by the multiple Hurwitz zeta function of order $2$ and the multiple gamma function of order $2$, respectively. Moreover, the functional equation is given by the multiple sine function of order $2$.
\par
Case (ii).
\par
\begin{align}
f_2 (x) = \frac{1}{(x -1)^2 (x^2-1)}.
\label{kirishima02}
\end{align}
Thus we find 
\begin{align*}
f_2 \left( \frac{1}{x} \right) = - x^{2^2} f_2(x),
\end{align*}
so $f_2$ is an absolute automorphic form of weight $-4 = -2^2$. Noting that $\ell =0, \ a=0, \ b=3, \ n(1)=n(2)=1, \ n(3)=2, \ {\rm deg} (f_2) = D = -2^2, \ \bec{n} =(1,1,2), \  C=-1,$ and $\varepsilon_{f_2} (s) = S_3 \left( s + 2^2, (1,1,2) \right)$, by Theorem 1, we get
\begin{align*}
Z_{f_2} (w, s) 
&= \zeta_3 \left( w, s + 2^2, (1,1,2) \right),
\\
\zeta_{f_2} (s)
&= \Gamma_3 \left( s + 2^2, (1,1,2) \right),
\\
\zeta_{f_2} (-2^2-s)^{-1}
&= S_3 \left( s + 2^2, (1,1,2) \right) \ \zeta_{f_2} (s).
\end{align*}
Therefore we see that $Z_{f_2} (w,s)$ and $\zeta_{f_2} (s)$ are given by the multiple Hurwitz zeta function of order $3$ and the multiple gamma function of order $3$, respectively. In addition, the functional equation is expressed by the multiple sine function of order $3$.
\par
Case (iii).
\par
\begin{align}
f_3 (x) = \frac{1}{(x^2 -1)^4}.
\label{kirishima02}
\end{align}
Then we have
\begin{align*}
f_3 \left( \frac{1}{x} \right) = x^{2^3} f_3(x),
\end{align*}
thus $f_3$ is an absolute automorphic form of weight $-2^3$. Noting that $\ell =0, \ a=0, \ b=4, \ n(1)=n(2)=n(3)=n(4)=2, \ {\rm deg} (f_3) = D = -2^3, \ \bec{n} =(2,2,2,2), \  C=1,$ and $\varepsilon_{f_3} (s) = S_4 \left( s + 2^3, (2,2,2,2) \right)$, from Theorem 1, we obtain
\begin{align*}
Z_{f_3} (w, s) 
&= \zeta_4 \left( w, s + 2^3, (2,2,2,2) \right),
\\
\zeta_{f_3} (s)
&= \Gamma_4 \left(s + 2^3, (2,2,2,2) \right),
\\
\zeta_{f_3} (-2^3-s)
&= S_4 \left( s + 2^3, (2,2,2,2) \right) \ \zeta_{f_3} (s).
\end{align*}
So $Z_{f_3} (w,s)$ and $\zeta_{f_3} (s)$ can be expressed by the multiple Hurwitz zeta function of order $4$ and the multiple gamma function of order $4$, respectively. Furthermore, the functional equation is given by the multiple sine function of order $4$.
\par
Case (iv).
\par
\begin{align}
f_4 (x) = \frac{(x-1)^4}{(x^2 -1)^{10}}.
\label{kirishima02}
\end{align}
Then we have
\begin{align*}
f_4 \left( \frac{1}{x} \right) = x^{2^4} f_4(x),
\end{align*}
therefore $f_4$ is an absolute automorphic form of weight $-2^4$. Noting that $\ell =0, \ a=4, \ m(1)= \cdots =m(4)=1, \ b=10, \ n(1)= \cdots =n(10)=2, \ {\rm deg} (f_4) = D = -2^4, \ \bec{n} = (\overbrace{2, \ldots, 2}^{10}), \  C=1,$ and 
\begin{align*}
\varepsilon_{f_4} (s) = 
\prod_{I \subset \{1, 2, 3, 4 \}} S_{10} \left( s +2^4 + |I|, \bec{n} \right)^{ (-1)^{|I|}},
\end{align*}
it follows from Theorem 1 that
\begin{align*}
Z_{f_4} (w, s) 
&= \sum_{I \subset \{1, 2, 3, 4 \}} (-1)^{|I|} \ \zeta_{10} \left(w, s +2^4 + |I|, \bec{n} \right),
\\
\zeta_{f_4} (s)
&= \prod_{I \subset \{1, 2, 3, 4 \}} \Gamma_{10} \left( s +2^4 + |I|, \bec{n} \right)^{ (-1)^{|I|}},
\\
\zeta_{f_4} (-2^4-s)
&= \left\{ \prod_{I \subset \{1, 2, 3, 4 \}} S_{10} \left( s +2^4 + |I|, \bec{n} \right)^{ (-1)^{|I|}} \right\} \ \zeta_{f_4} (s).
\end{align*}
Thus $Z_{f_4} (w,s)$ and $\zeta_{f_4} (s)$ can be obtained by the multiple Hurwitz zeta function of order $10$ and the multiple gamma function of order $10$, respectively. In addition, the functional equation is expressed by the multiple sine function of order $10$.

\section{QCA \label{sec03}}
This section gives the definition of our QCA. Let $\mathbb{P}_N = \{ 0,1, \ldots, N-1 \}$ be the path space with $N$ sites. Throughout this paper, we mainly assume that $N \ge 2$. There are two states ``0" or ``1" at each site. Let $\eta (x) \in \{0,1\}$ denote the state of the site $x \in \PM_N$, i.e., $x=0,1,\ldots, N-1$. The configuration space is $\{0,1\}^{\mathbb{P}_N}$ with $2^N$ elements. Intuitively, a configuration $\eta = (\eta (0), \eta (1), \ldots, \eta (N-1)) \in \{0,1\}^{\PM_N}$ is given an occupation interpretation as follows: $\eta (x)=1$ means that a particle exists at site $x \in \PM_N$, and $\eta (x)=0$ means that $x$ is vacant. In this paper, we put
\begin{align}
\miniket 0 =
\begin{bmatrix}
1 \\
0
\end{bmatrix} 
,
\quad 
\miniket 1 =
\begin{bmatrix}
0 \\
1
\end{bmatrix}
.
\label{zeroone} 
\end{align}
For example, when $N=3$, a configuration $(0,0,1) \in \{0,1\}^{\mathbb{P}_3}$ means that the state ``0" at site 0, the state ``0" at site 1, and the state ``1" at site 2. In other words, $(\eta (0), \eta (1), \eta (2)) = (0,0,1)$. We also write $(0,0,1)$ by $\miniket 0 \miniket 0 \miniket 1 = \miniket 0 \otimes \miniket 0 \otimes \miniket 1$, where $\otimes$ means the tensor product. By using Eq. \eqref{zeroone}, we have
\begin{align*}
\miniket 0 \miniket 0 \miniket 1 = 
\begin{bmatrix}
1 \\
0
\end{bmatrix} 
\otimes
\begin{bmatrix}
1 \\
0
\end{bmatrix} 
\otimes
\begin{bmatrix}
0 \\
1
\end{bmatrix} 
=
\begin{bmatrix}
0 \\
1 \\
0 \\
0 \\
0 \\
0 \\
0 \\
0 
\end{bmatrix}
\in \CM^{2^3}.
\end{align*}

To define our QCA, we introduce the {\em local} operator $Q^{(l)}$ and the {\em global} operator $Q^{(g)}_N$ in the following way. This definition is based on Katori et al. \cite{KatoriEtAl2004}.

We first define the $4 \times 4$ matrix $Q^{(l)}$ by
\begin{align*}
Q^{(l)}
=
\begin{bmatrix}
a^{00}_{00} & a^{01}_{00} & a^{10}_{00} & a^{11}_{00} \\ 
a^{00}_{01} & a^{01}_{01} & a^{10}_{01} & a^{11}_{01} \\ 
a^{00}_{10} & a^{01}_{10} & a^{10}_{10} & a^{11}_{10} \\ 
a^{00}_{11} & a^{01}_{11} & a^{10}_{11} & a^{11}_{11}  
\end{bmatrix} 
,
\end{align*}
where $a^{ij}_{kl} \in \mathbb{C}$ for $i,j,k,l \in \{0,1\}.$  Let $\eta_n (x) \in \{0,1\}$ denote the state of the site $x \in \PM_N$ at time $n \in \ZM_{\ge}$. The element of $Q^{(l)}$, $a^{ij}_{kl}$, means the {\em transition weight} from $(\eta_n (x), \eta_n (x+1)) =(i,j)$ to $(\eta_{n+1} (x), \eta_{n+1} (x+1))=(k,l)$ for any $x=0,1, \ldots, N-2$ and $n \in \ZM_{\ge}$. If $a^{ij}_{kl} \in [0,1]$, then the transition weight can be the transition probability. We call ``$x$" the {\em left site} and ``$x+1$" the {\em right site}. Throughout this paper, we assume that $a^{ij}_{kl}=0$ if $j \not=l$. In other words, after the time transition, the state of the rightmost site does not change. This assumption is necessary to define the global operator $Q^{(g)}_N$ described below. Therefore, under this assumption, $Q^{(l)}$ is rewritten as
\begin{align*}
Q^{(l)}
=
\begin{bmatrix}
a^{00}_{00} & \cdot & a^{10}_{00} & \cdot \\ 
\cdot & a^{01}_{01} & \cdot & a^{11}_{01} \\ 
a^{00}_{10} & \cdot & a^{10}_{10} & \cdot \\ 
\cdot & a^{01}_{11} & \cdot & a^{11}_{11}  
\end{bmatrix} 
,
\end{align*}
where ``$\cdot$" means 0. By definition, the interaction of our QCA is nearest neighbor. In particular, if $a^{ij}_{kl} \in \{0,1\}$, then the IPS is called the {\em cellular automaton} (CA). Next we define the $2^N \times 2^N$ matrix $Q^{(g)}_N$ by
\begin{align*}
Q^{(g)}_N
&= \left( I_2 \otimes I_2 \otimes \cdots \otimes I_2 \otimes Q^{(l)} \right) \left( I_2 \otimes I_2 \otimes \cdots \otimes Q^{(l)} \otimes I_2 \right) 
\\
& \qquad \cdots \left( I_2 \otimes Q^{(l)} \otimes \cdots \otimes I_2 \otimes I_2 \right) \left( Q^{(l)} \otimes I_2 \otimes \cdots \otimes I_2 \otimes I_2 \right),
\end{align*}
where $I_n$ is the $n \times n$ identity matrix. For example, if $N=3$, then the $2^3 \times 2^3$ matrix $Q^{(g)}_3$ is 
\begin{align*}
Q^{(g)}_3 
= \left( I_2 \otimes Q^{(l)} \right) \left( Q^{(l)} \otimes I_2 \right).
\end{align*}
If $N=4$, then the $2^4 \times 2^4$ matrix $Q^{(g)}_4$ is 
\begin{align*}
Q^{(g)}_4 
= \left( I_2 \otimes I_2 \otimes Q^{(l)} \right) \left( I_2 \otimes Q^{(l)} \otimes I_2 \right) \left( Q^{(l)} \otimes I_2 \otimes I_2 \right).
\end{align*}
Note that if $N=2$, then $Q^{(g)}_2 = Q^{(l)}$. Moreover, when $N=1$, we put $Q^{(g)}_1 = I_2$.

We see that when $N=4$, a transition weight from $(\eta_n (0), \eta_n (1), \eta_n (2), \eta_n (3)) =(i_0,i_1,i_2,i_3) \in \{0,1\}^4$ to $(\eta_{n+1} (0), \eta_{n+1} (1), \eta_{n+1} (2), \eta_{n+1} (3)) =(k_0,k_1,k_2,k_3) \in \{0,1\}^4$ is $a^{i_0 i_1}_{k_0 k_1} a^{i_1 i_2}_{k_1 k_2} a^{i_2 i_3}_{k_2 k_3}$ for any $n \in \ZM_{\ge}$, for instance.

The above mentioned model was called the {\em interacting particle systems} (IPS) in our previous paper \cite{KomatsuEtAl2022}. We considered two typical classes, one is QCA and the other is PCA. Note that PCA is also called stochastic CA. Our QCA satisfies that $Q^{(l)}$ is unitary, i.e., 
\begin{align*}
&
|a^{00}_{00}|^2 + |a^{00}_{10}|^2 = |a^{01}_{01}|^2 + |a^{01}_{11}|^2 = |a^{10}_{00}|^2 + |a^{10}_{10}|^2 = |a^{11}_{01}|^2 + |a^{11}_{11}|^2 = 1,
\nonumber
\\
& 
\qquad a^{00}_{00} \  \overline{a^{10}_{00}} + a^{00}_{10} \  \overline{a^{10}_{10}} =
 a^{01}_{01} \  \overline{a^{11}_{01}} + a^{01}_{11} \  \overline{a^{11}_{11}} =0. 
\end{align*}
This QCA was introduced by Konno \cite{Konno2008b} as a quantum counterpart of the Domany-Kinzel model \cite{DomanyKinzel1984} which is a typical model of PCA. we easily see that `` $Q^{(l)}$ is a unitary matrix if and only if $Q^{(g)}_N$ is a unitary matrix". Moreover, `` $Q^{(l)}$ is an orthogonal matrix if and only if $Q^{(g)}_N$ is an orthogonal matrix". 

On the other hand, a model in PCA satisfies 
\begin{align}
a^{00}_{00} + a^{00}_{10} = a^{01}_{01} + a^{01}_{11} = a^{10}_{00} + a^{10}_{10} = a^{11}_{01} + a^{11}_{11} = 1, \quad a^{ij}_{kj} \in [0,1].
\label{condPCA}
\end{align}
That is, $Q^{(l)}$ becomes a transposed {\em stochastic matrix} (also called {\em transition matrix}). Furthermore, as in the case of the QCA, we find that `` $Q^{(l)}$ is a transposed stochastic matrix if and only if $Q^{(g)}_N$ is a transposed stochastic matrix". In other words, the sum of the elements of any column for $Q^{(l)}$ or $Q^{(g)}_N$ is equal to 1.

The evolution of QCA on $\PM_N$ is determined by 
\begin{align*}
\eta_{n} = \left( Q^{(g)} _N \right)^n \eta_{0} \quad (n \in \ZM_{\ge})
\end{align*}
for an initial state $\eta_{0}$. Note that $\eta_{n}, \eta_{0} \in \CM^{2^N}$ and $Q^{(g)} _N$ is a $2^N \times 2^N$ matrix.

For example, when $N=3$ and $(\eta (0), \eta (1), \eta (2)) = (0,0,1) \in \CM^{2^3}$, we observe
\begin{align*}
Q^{(g)}_3 (0,0,1)
= a^{00}_{00} a^{01}_{01} (0,0,1) + a^{00}_{00} a^{01}_{11} (0,1,1) 
+ a^{00}_{10} a^{01}_{01} (1,0,1) + a^{00}_{10} a^{01}_{11} (1,1,1).  
\end{align*}
In the case of QCA, the probability that a configuration $(0,1,1)$ exists is $|a^{00}_{00} a^{01}_{11}|^2$. On the other hand, as for PCA, the probability that a configuration $(0,1,1)$ exists is $a^{00}_{00} a^{01}_{11}$.

\section{Zeta functions for QCA \label{sec04}}
In our previous paper \cite{KomatsuEtAl2022}, we defined the {\em IPS-type zeta function} by 
\begin{align}
\overline{\zeta} \left(Q^{(l)}, \PM_N, u \right) = \left\{ \det \Big( I_{2^N} - u Q^{(g)}_N \Big) \right\}^{-1/2^N} \quad (N \in \ZM_{>}),
\label{satosan01ips}
\end{align}
for $u \in \mathbb{R}$. Here we introduce our zeta function for QCA as follows:
\begin{align}
\zeta \left(Q^{(l)}, \PM_N, u \right) = \det \Big( I_{2^N} - u Q^{(g)}_N \Big)^{-1} \quad (N \in \ZM_{>}).
\label{kakumei02qca}
\end{align}
The relation between IPS-type and our zeta functions is 
\begin{align*}
\zeta \left(Q^{(l)}, \PM_N, u \right) = \overline{\zeta} \left(Q^{(l)}, \PM_N, u \right)^{2^N}. 
\end{align*}
Remark that when $N=1$, we put $Q^{(g)}_1 = I_2$. Thus we find 
\begin{align}
\zeta \left(Q^{(l)}, \PM_1, u \right) = \overline{\zeta} \left(Q^{(l)}, \PM_1, u \right)^2 = (1-u)^{-2}.
\label{zetaN1}
\end{align}
In order to clarify the dependence on $N$, from now on, we define ``$\zeta \left(Q^{(l)}, \PM_N, u \right)$" by ``$\zeta_N (u)$". So Eq. \eqref{kakumei02qca} is rewritten as 
\begin{align}  
\zeta_N (u) = \left\{ \det \left( I_{2^N} - u Q^{(g)}_N \right) \right\}^{-1}.
\label{kakumei03}
\end{align}

Here, we consider a zeta function $\zeta_{A} (u)$ for a general $M \times M$ matrix $A$ defined by
\begin{align}  
\zeta_{A} (u) = \left\{ \det \left(I_{M} - u  A \right) \right\}^{-1}.
\label{mushiastui01}
\end{align}
If $A$ is a regular matrix with its eigenvalues $\{\lambda_1, \ldots, \lambda_M \}$, then we easily find
\begin{align*}  
\zeta_{A} \left( \frac{1}{u} \right)^{-1} 
&= \det \left( I_{M} - \frac{1}{u} A \right)
= \prod_{k=1}^M \left( 1 - \frac{\lambda_k}{u} \right)
= \left(\frac{1}{u} \right)^M \prod_{k=1}^M \left( u - \lambda_k \right)
\\
&
= \left(\frac{1}{u} \right)^M \left( \prod_{k=1}^M \lambda_k \right) \ (-1)^M \ \prod_{k=1}^M \left( 1 - \frac{u}{\lambda_k} \right)
\\
&
= (-u)^{-M} \ \det A \ \left\{ \zeta_{{A}^{-1}} \left( u \right) \right\}^{-1}. 
\end{align*}
Therefore we obtain the following result.
\begin{align}
\zeta_{A} \left( \frac{1}{u} \right) = (-u)^{M} \ \left(\det A \right)^{-1} \ \zeta_{A^{-1}} \left( u \right).
\label{kakumei01}
\end{align}
If $A$ is an orthogonal matrix, then 
\begin{align} 
A^{-1} = {}^{\rm{T}} \! \! A,
\label{kakumei05qca}
\end{align}
where $\rm{T}$ is the transposed operator. From Eq. \eqref{kakumei05qca}, we find
\begin{align*}  
\zeta_A (u) 
&= \left\{ \det \left( I_{M} - u A \right) \right\}^{-1} = \left\{ \det \left( I_{M} - u {}^{\rm{T}} \! \! A \right) \right\}^{-1} 
\\
&= \left\{ \det \left( I_{M} - u A^{-1} \right) \right\}^{-1} =\zeta_{A^{-1}} (u). 
\end{align*}
That is, 
\begin{align}
\zeta_A (u) = \zeta_{A^{-1}} (u).  
\label{kakumei05bqca}
\end{align}
Moreover, if $A$ is an orthogonal matrix, then
\begin{align}
\left(\det A \right)^{-1} = \det A,
\label{kakumei05cqca}
\end{align}
since $\det A \in \{-1,1\}$. Combining Eq. \label{kakumei01} with Eqs. \eqref{kakumei05qca}, \eqref{kakumei05bqca} and \eqref{kakumei05cqca} implies that if $A$ is an orthogonal matrix, then
\begin{align}
\zeta_{A} \left( \frac{1}{u} \right) = (-u)^{M} \ \det A \ \zeta_{A} \left( u \right).
\label{kakumei01bqca}
\end{align}
Then if $Q^{(g)}_N$ is orthogonal, then it follows from Eq. \eqref{kakumei01bqca} with $M=2^N$ that
\begin{align*}
\zeta_{N} \left( \frac{1}{u} \right) = u^{2^N} \ \left(\det Q^{(g)}_N \right) \ \zeta_{N} \left( u \right).
\end{align*}
Therefore we obtain the following result for our zeta function $\zeta_{N} (u)$.

\vspace*{12pt}
\noindent
{\bf Theorem~2:} 
If $Q^{(g)}_N$ is an orthogonal matrix, then we have
\begin{align}
\zeta_{N} \left( \frac{1}{u} \right) = \det Q^{(g)}_N \ u^{2^N} \ \zeta_{N} \left( u \right),
\label{kakumei08qca}
\end{align}
where $\det Q^{(g)}_N \in \{-1,1\}$. 

\vspace*{12pt}
Recall that $f$ is an absolute automorphic form of weight $D$ if $f$ satisfies
\begin{align*}
f \left( \frac{1}{u} \right) = C \ u^{-D} \ f(u)
\end{align*}
with $C \in \{ -1, 1 \}$ and $D \in \mathbb{Z}$. Therefore, from Theorem 2, we have an important property of our zeta function $\zeta_{N} (u)$, that is, ``$\zeta_{N} (u)$ is an absolute automorphic form of weight $- 2^N$". Then $\zeta_{\zeta_{N}} (s)$ is a absolute zeta function for our zeta function $\zeta_{N} (u)$. In other words, we can consider ``the zeta function of a zeta function".

In our previous paper \cite{KomatsuEtAl2022}, we studied two classes for QCA. One is given by
\begin{align*}
Q^{(l)}_{QCA,1} (\xi_1, \xi_2)
=
\begin{bmatrix}
\cos \xi_1 & \cdot & - \sin \xi_1 & \cdot \\ 
\cdot & \cos \xi_2 & \cdot & - \sin \xi_2 \\ 
\sin \xi_1 & \cdot & \cos \xi_1 & \cdot \\ 
\cdot & \sin \xi_2 & \cdot & \cos \xi_2  
\end{bmatrix} 
,
\end{align*}
where $\xi_1, \xi_2 \in [0, 2 \pi)$. In particular, $Q^{(l)}_{QCA,1} (0,0) = I_4$. The other is 
\begin{align*}
Q^{(l)}_{QCA,2} (\xi_1, \xi_2)
=
\begin{bmatrix}
\cos \xi_1 & \cdot & - \sin \xi_1 & \cdot \\ 
\cdot & - \sin \xi_2 & \cdot & \cos \xi_2 \\ 
\sin \xi_1 & \cdot & \cos \xi_1 & \cdot \\ 
\cdot & \cos \xi_2 & \cdot & \sin \xi_2  
\end{bmatrix} 
,
\end{align*}
where $\xi_1, \xi_2 \in [0, 2 \pi)$. In particular, $Q^{(l)}_{QCA,2} (0,0)$ becomes the well-known Wolfram {\em Rule 90} (see \cite{Durett1988}, for example). It is noted that Rule 90 is not only QCA but also PCA. 

We should remark that $Q^{(l)}_{QCA,1} (\xi_1, \xi_2)$ and $Q^{(l)}_{QCA,2} (\xi_1, \xi_2)$ are orthogonal matrices, so Theorem 2 holds for both models. In other words, ``$\zeta_{N} (u)$ defined by $Q^{(l)}_{QCA,1} (\xi_1, \xi_2)$ or $Q^{(l)}_{QCA,2} (\xi_1, \xi_2)$ is an absolute automorphic form of weight $- 2^N$".

\section{Examples \label{sec05}}
In this section, we consider a tensor-type model with $Q^{(l)} = Q^{(l)}_{QCA,2} (0, \xi)$ as follows:
\begin{align}
Q^{(l)}_{QCA,2} (0, \xi)
=
I_2 \otimes E_{00} + 
\sigma (\xi)
\otimes
E_{11}
\qquad (\xi \in [0, 2 \pi)),
\label{oto01}
\end{align}
where 
\begin{align*}
\sigma (\xi) = 
\begin{bmatrix}
- \sin \xi & \cos \xi \\
\cos \xi & \sin \xi 
\end{bmatrix}
, \quad
E_{00} =
\begin{bmatrix}
1 & 0 \\
0 & 0 
\end{bmatrix}
, \quad
E_{11} =
\begin{bmatrix}
0 & 0 \\
0 & 1 
\end{bmatrix}
.
\end{align*}
Note that
\begin{align*}
\sigma (\xi)^2 = I_2 \qquad (\xi \in [0, 2 \pi)).
\end{align*}
From now on, we put 
\begin{align*}
Q^{(l)} (\xi) = Q^{(l)}_{QCA,2} (0, \xi).
\end{align*}
Thus we see
\begin{align*}
Q^{(l)} (\xi) =
\begin{bmatrix}
1 & \cdot & \cdot & \cdot \\ 
\cdot & - \sin \xi & \cdot & \cos \xi \\ 
\cdot & \cdot & 1 & \cdot \\ 
\cdot & \cos \xi & \cdot & \sin \xi  
\end{bmatrix} 
.
\end{align*}

Let ${\rm Spec} (A)$ be the set of eigenvalues of a square matrix $A$. More precisely, we also use the following notation: 
\begin{align*}
{\rm Spec} (A) = \left\{ \left[ \lambda_1 \right]^{l_1}, \ \left[ \lambda_2 \right]^{l_2}, \ \ldots \ , \left[ \lambda_k \right]^{l_k} \right\},
\end{align*}
where $\lambda_j$ is the eigenvalue of $A$ and $l_j \in \ZM_{>}$ is the multiplicity of $\lambda_j$ for $j=1,2, \ldots, k$. Let $\{ T_n (x) \}$ denote the Chebychev polynomials of the first kind (see Andrews et al. \cite{Andrews1999}):
\begin{align*}
T_n (x) = \cos \left( n \cdot \cos^{-1} (x) \right) \qquad \left(n \in \ZM_{\ge}, \ x \in [-1,1] \right),
\end{align*}
where $\cos^{-1} (x) = {\rm arccos} (x)$. For example, 
\begin{align*}
T_0 (x) = 1, \quad T_1 (x) = x, \quad T_2 (x) = 2x^2-1, \quad T_3 (x) = 4x^3-3x, \ldots.
\end{align*}
Then, by using Proposition 5 (i) for $r=1$ case in \cite{KomatsuEtAl2022}, we have the following result. 

\vspace*{12pt}
\noindent
{\bf Proposition~3:} 
\begin{align*}
{\rm Spec} \left( Q^{(g)}_N \left( \pi/2 \right)\right) = \left\{ \left[ 1 \right]^{c_N (1)}, \ \left[ -1 \right]^{c_N(-1)} \right\},
\end{align*}
where 
\begin{align*}
c_N (1) = \frac{1}{2} \left(2^N + B_N \right), \quad c_N (-1) = \frac{1}{2} \left(2^N - B_N \right), \quad B_N = 2^{(N+1)/2} \  T_{N-1} \left( \sqrt{2}/2 \right).
\end{align*}

\vspace*{12pt}
Remark that $c_N (1) + c_N (-1) = 2^N.$ From now on, we put 
\begin{align*}
Q^{(g)}_N = Q^{(g)}_N \left( \pi/2 \right).
\end{align*}
Therefore, by Eq. \eqref{kakumei03} and Proposition 3, we see
\begin{align*}  
\zeta_N (u) 
&= \left\{ \det \left( I_{2^N} - u Q^{(g)}_N \right) \right\}^{-1} 
= \prod_{j=1}^{2^N} \frac{1}{1-\lambda_j u}
= \left( \frac{1}{1-u} \right)^{c_N (1)} \left( \frac{1}{1+u} \right)^{c_N (-1)}\\
&= (-1)^{c_N (1)} g_N(u),
\end{align*}
where $\{ \lambda_j : j=1, \ldots, 2^N \}$ is the eigenvalue of $Q^{(g)}_N$ and 
\begin{align} 
g_N (u) = \frac{(u-1)^{c_N (-1)}}{(u-1)^{c_N (1)}(u^2-1)^{c_N (-1)}}.
\label{qca001}
\end{align}
Then we have
\begin{align*}
g_N \left( \frac{1}{u} \right) = (-1)^{c_N(1)} u^{2^N} g_N (u),
\end{align*}
therefore $g_N$ is an absolute automorphic form of weight $-2^N$.

We should note that $g_N (u) = f_N (u)$ for $(N=1,2,3,4)$, where $f_N (u)$ was defined in Section \ref{sec02}.

From now on, we consider three cases, i.e., Case (a): $c_N (1) > c_N (-1)$, Case (b): $c_N (1) = c_N (-1)$, and Case (c): $c_N (1) < c_N (-1)$.  
\par
\
\par
Case (a): $c_N (1) > c_N (-1)$. In this case, Eq. \eqref{qca001} becomes 
\par
\begin{align} 
g_N (u) = \frac{1}{(u-1)^{c_N (1)-c_N (-1)}(u^2-1)^{c_N (-1)}}.
\label{casea}
\end{align}
Noting that $\ell =0, \ a=0, \ b=c_N (1), \ n(1)= \cdots =n(c_N (1)-c_N (-1))=1, \ n(c_N (1)-c_N (-1)+1)= \cdots =n(c_N (1))=2, \ {\rm deg} (g_N) = D = -2^N, \  C=1,$ and 
\begin{align*}
\bec{n} 
&= \overbrace{(1, \ldots, 1,}^{c_N (1)-c_N (-1)} \overbrace{2, \ldots, 2}^{c_N (-1)} ), \qquad 
\varepsilon_{g_N} (s) 
= 
S_{2^N} \left( s +2^N, \bec{n} \right),
\end{align*}
it follows from Theorem 1 that
\begin{align*}
Z_{g_N} (w, s) 
&= \zeta_{2^N} \left(w, s +2^N, \bec{n} \right),
\\
\zeta_{g_N} (s)
&= \Gamma_{2^N} \left( s +2^N, \bec{n} \right),
\\
\zeta_{g_N} (-2^N-s)
&= S_{2^N} \left( s +2^N, \bec{n} \right) \ \zeta_{g_N} (s).
\end{align*}
We see that Case (i) in Section \ref{sec02} is Case (a) for $N=1$, since we find that $c_1 (1) = 2 > 0 = c_1 (-1), \ c_1 (1) - c_1 (-1) = 2, \ \ell =0, \ a=0, \ b=2^1, \ n(1)= n(2)=1, \ \bec{n} = (1,1), \ {\rm deg} (g_1) = D = -2^1$, and $C=1$. In addition, we confirm that Case (ii) in Section \ref{sec02} is Case (a) for $N=2$, because we see that $c_2 (1) = 3 > 1 = c_2 (-1), \ c_2 (1) - c_2 (-1) = 2, \ \ell =0, \ a=0, \ b=3, \ n(1)= n(2)=1, \ n(3)=2, \ \bec{n} = (1,1,2), \ {\rm deg} (g_2) = D = -2^2$, and $C=-1$.
\par
\
\par
Case (b): $c_N (1) = c_N (-1)$. In this case, Eq. \eqref{qca001} becomes 
\par
\begin{align} 
g_N (u) = \frac{1}{(u^2-1)^{c_N (-1)}}.
\label{casea}
\end{align}
Noting that $\ell =0, \ a=0, \ b=c_N (1)=2^{N-1}, \ n(1)= \cdots =n(2^{N-1})=2,  \ {\rm deg} \ g_N = D = -2^N, \  C=1,$ and 
\begin{align*}
\bec{n} 
&= ( \overbrace{2, \ldots, 2}^{c_N (-1)} ), \qquad 
\varepsilon_{g_N} (s) 
= 
S_{2^N} \left( s +2^N, \bec{n} \right),
\end{align*}
it follows from Theorem 1 that
\begin{align*}
Z_{g_N} (w, s) 
&= \zeta_{2^N} \left(w, s +2^N, \bec{n} \right),
\\
\zeta_{g_N} (s)
&= \Gamma_{2^N} \left( s +2^N, \bec{n} \right),
\\
\zeta_{g_N} (-2^N-s)
&= S_{2^N} \left( s +2^N, \bec{n} \right) \ \zeta_{g_N} (s).
\end{align*}
We see that Case (iii) in Section \ref{sec02} is Case (b) for $N=3$, since we find that $c_3 (1) = 4 = c_3 (-1), \ c_3 (1) - c_3 (-1) = 0, \ \ell =0, \ a=0, \ b=4, \ n(1)= n(2)=n(3)=n(4)=2, \ \bec{n} = (2,2,2,2), \ {\rm deg} (g_3) = D = -2^3$, and $C=1$. 
\par
\
\par
Case (c): $c_N (1) < c_N (-1)$. In this case, Eq. \eqref{qca001} becomes 
\par
\begin{align} 
g_N (u) = \frac{(u-1)^{c_N (-1)-c_N (1)}}{(u^2-1)^{c_N (-1)}}.
\label{casea}
\end{align}
Noting that $\ell =0, \ a=c_N (-1)-c_N (1), \ b=c_N (-1), \ m(1)= \cdots =m(c_N (-1)-c_N (1))=1, \ n(1)= \cdots =n(c_N (-1))=2, \ {\rm deg} (g_N) = D = -2^N, \  C=1,$ and 
\begin{align*}
\bec{n} 
&= (\overbrace{2, \ldots, 2}^{c_N (-1)} ),
\\
\varepsilon_{g_N} (s) 
&= 
\prod_{I \subset \{1, \ldots, c_N (-1)-c_N (1) \}} S_{c_N (-1)} \left( s +2^N + |I|, \bec{n} \right)^{ (-1)^{|I|}},
\end{align*}
it follows from Theorem 1 that
\begin{align*}
Z_{g_N} (w, s) 
&= \sum_{I \subset \{1, \ldots, c_N (-1)-c_N (1) \}} (-1)^{|I|} \ \zeta_{c_N (-1)} \left(w, s +2^N + |I|, \bec{n} \right),
\\
\zeta_{g_N} (s)
&= \prod_{I \subset \{1, \ldots, c_N (-1)-c_N (1) \}} \Gamma_{c_N (-1)} \left( s +2^N + |I|, \bec{n} \right)^{ (-1)^{|I|}},
\\
\zeta_{g_N} (-2^N-s)
&= \left\{ \prod_{I \subset \{1, \ldots, c_N (-1)-c_N (1) \}} S_{c_N (-1)} \left( s +2^N + |I|, \bec{n} \right)^{ (-1)^{|I|}} \right\} \ \zeta_{g_N} (s).
\end{align*}
We see that Case (iv) in Section \ref{sec02} is Case (c) for $N=4$, since we find that $c_4 (1) = 6 < 10 = c_4 (-1), \ \ell =0, \ a=c_4 (-1) - c_4 (1)=4, \ b=c_4 (-1)=10, \ m(1)= \cdots =m(4)=1, \ n(1)= \cdots =n(10)=2, \ {\rm deg} (f_4) = D = -2^4, \ \bec{n} = (\overbrace{2, \ldots, 2}^{10})$ and $C=1$. 
\par
\
\par
Hence, noting $\zeta_N (u) = (-1)^{c_N (1)} g_N(u)$, by results of Cases (a), (b) and (c) for $g_N(u)$, we obtain the following result for a tensor-type QCA.

\vspace*{12pt}
\noindent
{\bf Theorem~4:}
For Case (a): $c_N (1) > c_N (-1)$, we have
\begin{align*}
Z_{\zeta_{N}} (w, s) 
&= (-1)^{c_N (1)} \zeta_{2^N} \left(w, s +2^N, \bec{n} \right),
\\
\zeta_{\zeta_N} (s)
&= \Gamma_{2^N} \left( s +2^N, \bec{n} \right)^{(-1)^{c_N (1)}},
\\
\zeta_{\zeta_N} (-2^N-s)
&= S_{2^N} \left( s +2^N, \bec{n} \right)^{(-1)^{c_N (1)}} \ \zeta_{\zeta_N} (s).
\end{align*}
For Case (b): $c_N (1) = c_N (-1)$, we have
\begin{align*}
Z_{\zeta_{N}} (w, s) 
&= (-1)^{c_N (1)} \zeta_{2^N} \left(w, s +2^N, \bec{n} \right),
\\
\zeta_{\zeta_N} (s)
&= \Gamma_{2^N} \left( s +2^N, \bec{n} \right)^{(-1)^{c_N (1)}},
\\
\zeta_{\zeta_N} (-2^N-s)
&= S_{2^N} \left( s +2^N, \bec{n} \right)^{(-1)^{c_N (1)}} \ \zeta_{\zeta_N} (s).
\end{align*}
For Case (c): $c_N (1) < c_N (-1)$, we have
\begin{align*}
Z_{\zeta_{N}} (w, s) 
&= (-1)^{c_N (1)} \sum_{I \subset \{1, \ldots, c_N (-1)-c_N (1) \}} (-1)^{|I|} \ \zeta_{c_N (-1)} \left(w, s +2^N + |I|, \bec{n} \right),
\\
\zeta_{\zeta_N} (s)
&= \prod_{I \subset \{1, \ldots, c_N (-1)-c_N (1) \}} \Gamma_{c_N (-1)} \left( s +2^N + |I|, \bec{n} \right)^{ (-1)^{|I|+c_N (1)}},
\\
\zeta_{\zeta_N} (-2^N-s)
&= \left\{ \prod_{I \subset \{1, \ldots, c_N (-1)-c_N (1) \}} S_{c_N (-1)} \left( s +2^N + |I|, \bec{n} \right)^{ (-1)^{|I||+c_N (1)}} \right\} \ \zeta_{\zeta_N} (s).
\end{align*}

\vspace*{12pt}
We should remark that $Z_{\zeta_N} (w, s)$ and $\zeta_{\zeta_N} (s)$ can be obtained by the multiple Hurwitz zeta function of order $2^N$ (Cases (a) and (b)) or  $c_N (-1)$ (Case (c)) and the multiple gamma function of order $2^N$ (Cases (a) and (b)) or  $c_N (-1)$ (Case (c)), respectively. In addition, the functional equation is expressed by the multiple sine function of order $2^N$ (Cases (a) and (b)) or  $c_N (-1)$ (Case (c)).

\section{Summary \label{sec06}}
In this paper, we introduced our zeta function $\zeta_N (u)$ determined by a $Q^{(g)}_N$ which is a $2^N \times 2^N$ time evolution matrix of QCA on the path space $\mathbb{P}_N$. Then we proved that if $Q^{(g)}_N$ is an orthogonal matrix, then $\zeta_N (u)$ is an absolute automorphic form of weight $- 2^N$ (Theorem 2). After that we considered an absolute zeta function $\zeta_{\zeta_N} (s)$ for our zeta function $\zeta_N (u)$. As an example, we computed $\zeta_{\zeta_N} (s)$ for a tensor-type QCA, and showed that it is expressed as the multiple gamma function via the multiple Hurwitz zeta function (Theorem 4). Moreover, we obtained the functional equation for $\zeta_{\zeta_N} (s)$ by the multiple sine function (Theorem 4). The present manuscript is the first step of the study on a relationship between the QCA and the absolute zeta function. One of the interesting future problems might be to extend QCA on the one-dimensional path space to QCA on the higher-dimensional lattice.


\end{document}